# On the absence of structure factors in concentrated colloidal suspensions and nanocomposites

Anne-Caroline Genix and Julian Oberdisse *

*Laboratoire Charles Coulomb (L2C), Université de Montpellier, CNRS, F-34095 Montpellier, France*

* Corresponding author: julian.oberdisse@umontpellier.fr

**Abstract**

Small-angle scattering is a commonly used tool to analyze the dispersion of nanoparticles in all kinds of matrices. Besides some obvious cases, the associated structure factor is often complex and cannot be reduced to a simple interparticle interaction, like excluded volume only. In recent experiments, we have encountered a surprising absence of structure factors ($S(q) = 1$) in scattering from rather concentrated polymer nanocomposites [A.-C. Genix et al, ACS Appl. Mater. Interfaces 11 (2019) 17863]. In this case, quite pure form factor scattering is observed. This somewhat "ideal" structure is further investigated here making use of reverse Monte Carlo simulations in order to shed light on the corresponding nanoparticle structure in space. By fixing the target "experimental" apparent structure factor to one over a given q-range in these simulations, we show that it is possible to find dispersions with this property. The influence of nanoparticle volume fraction and polydispersity has been investigated, and it was found that for high concentrations only a high polydispersity allows reaching a state of $S = 1$. The underlying structure in real space is discussed in terms of the pair-correlation function, which evidences the importance of attractive interactions between polydisperse nanoparticles. The calculation of partial structure factors shows that there is no specific ordering of large or small particles, but that the presence of attractive interactions together with polydispersity allows reaching an almost "structureless" state.

**Graphical Abstract:**

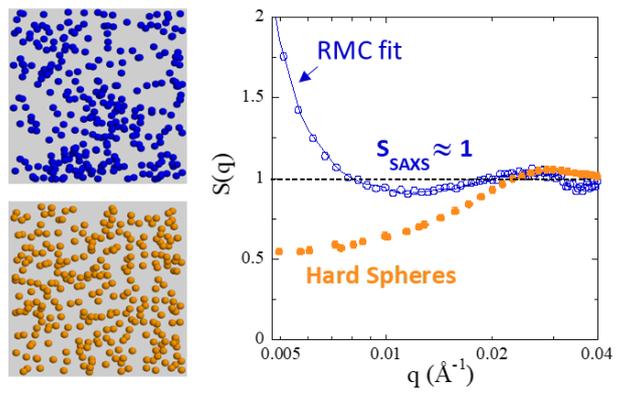

1. **Introduction**

Small-angle scattering of neutrons (SANS) and X-rays (SAXS) has undergone a strong instrumental development over the past decades, starting from close to zero in the 1960s. This has generated considerable benefits in terms of higher flux and faster data acquisition, more extended q-ranges, and innovative sample environments. Concerning in particular neutron scattering, the implementation of what is probably the best-performing SANS-suite of instruments (Dnn, n=1,2,3) in the world at ILL in Europe has undoubtedly contributed to the success of many scientific projects, thereby opening new fields of study. We can only comment on soft matter science, which however has enormously benefitted from ILL's SANS beamlines. Clearly, the importance of H-D contrast in soft matter, combined with accessible energy ranges in QENS, makes neutrons a unique tool, which however necessitates a complete working environment which was not obvious some 50 years ago. Just to cite one example, it has often been recalled that detectors had to be moved by hand in the infancy of SANS, and then detector tubes to be pumped empty again and again. Nowadays, one just presses a button, and the next measurement at a different q-range can start minutes later.

On a personal note, one of the authors (JO) started small-angle scattering with a neutron experiment on D22 with Roland May, quite exactly 30 years ago, together with his PhD supervisor Grégoire Porte, from University of Montpellier, before continuing for many years on the PA-suite at LLB (Saclay), and then returning to D11. Among the many good memories he has, the 1996 Bombannes school on scattering methods applied to soft condensed matter, organized by ILL, and in particular by Thomas Zemb and Peter Lindner, with Roland May participating as teacher, was simply outstanding. He is happy that in spite of the existence of modern teaching tools (which did not exist back then), the "Bombannes spirit" still continues, and we celebrated 30 years of Bombannes at the last edition in 2022, in presence of both Thomas and (the still organizer!) Peter. Coming back to D11, the full extent of technical developments is due to many people, among which the instrument scientists and technicians. Although we did not encounter all of them personally, it is known that Konrad Ibel was the historical driving force. Much later, when we did our first experiments, we had the pleasure to work with Peter Lindner, Johannes Zipfel, Ralf Schweins, and David Bowyer.

Besides preparing possibly complex samples with H-D contrasts, obtaining beamtime and doing the measurements, analysis of small-angle scattering (SAS) remains a challenge in many systems. This is due to the absence of any one-on-one correspondence between structure and scattering. There is hope, however, in the existence of more general statistical properties describing sample morphology on the nanoscale, like the averages of sizes, masses, volumes, surfaces, or distances – we usually term them "key features" of a sample. To illustrate this, one may imagine any given configuration of nanoparticles in space. They give a response in elastic scattering, as e.g., obtained by small-angle scattering from particle assemblies. The loss in phase information and the limited survey of q-space makes the scattered intensity an incomplete representation of the real particle assembly. This is particularly obvious when the number of particles in a sample, each with a position defined by its coordinates in space, exceeds the number of points in q-space by many orders of magnitude. Nonetheless, it is possible to obtain a statistically relevant set of configurations of particles the scattering of which is compatible with the measured intensity – these are simply the ones possessing the same (robust) key features as the real sample, as mentioned above. Reverse Monte Carlo (RMC) simulations are one pathway to obtaining such configurations; others, based on more mathematical methods, exist [1, 2]. The key advantage of RMC is here that an important part of the sample, namely the shape and size of the nanoparticles, is included as starting point of the analysis. In a first application of the RMC technique – initially developed for atomic structures [3-5] – individual aggregates of nanoparticles in polymer nanocomposites (PNCs) have been modelled [6]. In an attempt to describe anisotropic scattering of deformed samples, Hagita has modeled very large assemblies using supercomputers and now GPUs. [7-11] A few years ago, we have proposed a multi-particle model working with typically a few thousand particles in a cubic simulation box, capable of identifying percolation across the box [12].

In this article, we will discuss a problem of data analysis which we encountered recently in the framework of our nanocomposite studies [13]. All experimental data and procedures are taken from this reference; the present paper focuses on data analysis. In ref. [13], measurements by SAXS and SANS have been combined to highlight polymer structure around nanoparticles, and the nanoparticle structure within the polymer matrix. Our recently developed multi-particle simulation will be shown here to be the appropriate tool to analyze the surprising structural properties of polymer nanocomposites with pre-adsorbed polymer chains on the particles, where the structure factor was found to approach one at relatively high concentrations in non-annealed samples. The observed intensity thus superimposes quite well with the form factor measured independently at high dilution. Given that any user of the Percus-Yevick monodisperse hard-sphere structure factor knows that as soon as the particle concentration exceeds a few percent, there is a growing peak preceded by a deeper and deeper correlation hole in q-space, the superposition of the independently measured particle form factor, and the particle scattering at high concentration (≈ 10%v) is highly unexpected. We thus try to elucidate this mysterious structure using RMC here, or at least contribute to its understanding. Coming back to the importance of SANS instruments and in particular D11 for soft matter studies, it is noted that the nanoparticle data presented here could have been measured with neutrons, but X-rays were sufficient. This is not the case as soon as one is simultaneously interested in polymer structure, as explored by us with SANS in many different projects [13-16].

2. Methods

**Experimental system:** For convenience of the reader, experimental procedures used to produce the samples of poly(vinyl acetate)-silica-nanocomposites are shortly recalled from ref [13].

**Silica nanoparticles.** Spherical silica NPs were synthesized by a modified Stöber method in ethanol and they have been characterized by SAXS in ethanol at 0.7 vol % dilution. The form factor fits shown in this paper correspond to a log-normal size distribution with $R_0$ = 9.6 nm and polydispersity 18%, yielding an average NP radius a bit below 10 nm. Sometimes a different batch has been used, with equivalent characteristics.

**Polymer chains.** Most poly(vinyl acetate) (PVAc) has been purpose-synthesized and the molecular weights were characterized by size-exclusion chromatography in THF. The polymers used here are termed D10 for perdeuterated PVAc ($M_w$ = 10.6 kg/mol, PI = 1.2, $R_g$ = 2.7 nm in melt), H10 for hydrogenated PVAc ($M_w$ = 13.3 kg/mol, PI = 1.1, $R_g$ = 3.0 nm), and H40 (purchased from Alfa Aesar) for hydrogenated PVAc ($M_w$ = 41.0 kg/mol, PI = 2.4, $R_g$ = 5.2 nm). The deuteration protocol was used in order to highlight the chain scattering in ref [13]. With SAXS, there is no difference in contrast, and only NP scattering is measured, regardless of deuteration.

**Nanocomposite preparation.** The structure factor effect discussed in this paper is related to pre-adsorption of PVAc chains on the NPs. PVAc layers have been pre-adsorbed by mixing and stirring appropriate amounts of polymer in methyl ether ketone (MEK, ACS-grade, BDH VWR) with silica nanoparticles suspended in ethanol for 12 hours. This was followed by a triple centrifugation-redispersion in MEK. In this article, only H40(25%) pre-adsorbed samples are discussed and compared to PNCs without pre-adsorption ("bare NPs"). The former samples correspond to pre-adsorption of H40 as indicated above, for a total amount of 25%wt of polymer of the total pre-adsorbed object (particle and polymer). This amount of pre-adsorption can be expressed as an estimated dry thickness of the polymer layer (assuming a polymer density of 1.25 g/cm$^3$) of 1.2 nm. Polymer nanocomposites have then been formulated by dissolving the matrix polymer (D10, or D10/H10 blends, or H40/D10 blends) in MEK, and mix this solution (12 hours) with either bare or pre-adsorbed NPs in ethanol in appropriate proportions, aiming at a silica volume fraction $\Phi_{NP}$ of ca. 9%v in the final PNC obtained by solvent evaporation (fast: instantaneous, under vacuum; slow: progressive casting over one day). Exact NP volume fractions have been measured by TGA and are reported with the SAXS measurements.

**Reverse Monte Carlo simulations of particle assemblies probed by SAS**

RMC simulations have been described in previous publications [12, 17]. Spherical particles obeying the experimental size distribution of log-normal parameters ($R_0$, σ) are placed in a cubic simulation box with periodic boundary conditions, of size $L_{box}$ = 2π/$q_{min}$, where $q_{min}$ is the lowest experimentally accessible scattering wave vector. The number N of polydisperse particles sets the total particle volume, and thus the particle volume fraction, which is chosen to be equal to the experimental one (or systematically varied in the model calculations shown below). For theoretical calculations, $L_{box}$ has been fixed such that for each volume fraction the number of particles provided a satisfactory statistic of the results. In practice, this means studying larger boxes at lower concentrations, in order to have a few thousand particles in a box. The initial configuration was chosen to be random without collisions because particles possess excluded volume. Then, the Monte Carlo simulation proceeds by steps which are given by random moves of randomly chosen particles in the box, while respecting all excluded volumes. After each move, the scattered intensity (taking into account the polydispersity) is recalculated using the classical Debye expression [18] for the differential scattering cross section per unit sample volume:

$$I(q) = \frac{\Delta\rho^2}{V} \sum_{i,j=1}^{N} V_i V_j F_i(q) F_j(q) \frac{\sin(q d_{ij})}{q d_{ij}} \qquad (1)$$

where each $F_i(q)$ corresponds to the form factor amplitude of a given nanoparticle i (with its individual radius $R_i$), $d_{ij}$ is the center-to-center distance between two particles i and j, and q is the norm of the scattering vector. V is the simulation box volume. Low-q cross sections are determined by a lattice calculation as further discussed below [12, 17, 19-21]. For monodisperse particles of spherical symmetry eq.(1) reduces to the common factorization in form and structure factor. The scattering contrast Δρ is defined by the difference in scattering length density between the particles and the surrounding matrix. Its square is also the prefactor of the average form factor of the particles P(q), which is defined as the intensity scattered by the same particles in absence of interaction:

$$P(q) = \frac{\Delta\rho^2}{V} \sum_{i=1}^{N} V_i^2 F_i^2(q) \qquad (2)$$

The contrast squared cancels in the ratio of intensity I(q) to form factor P(q), and this ratio is called the apparent structure factor S(q). This structure factor is usually named "apparent" because, in presence of polydispersity, it is not the Fourier transform of the pair-correlation function of the centers of mass. Only for monodisperse assemblies, a single "true" structure factor exists, with its thermodynamic meaning. For simplicity, we drop the adjective "apparent" in this work on most occasions, and simply name this function S(q).

In order to obtain the correct high-q limit of the structure factor, the form factor measured at high dilution is usually superimposed to the high-q intensity, which amounts to rescaling it to the same contrast and concentration as the measured nanocomposite sample. This is automatically the case when using eq.(2). The calculated structure factor is then compared to the experimental one using the following definition of $\chi^2$:

$$\chi^2 = \frac{1}{N_q} \sum_{i=1}^{N_q} \left( \frac{S_{exp}(i) - S(i)}{\Delta S} \right)^2 \quad (3)$$

where $N_q$ is the number of points in q-space. Note that we have chosen error bars in this definition in accordance with the purpose outlined below, i.e., a comparison to a theoretical target function $S_{exp}(q)$ = 1. $\Delta S$ was set arbitrarily to 0.1. Using a simulated annealing algorithm, $\chi^2$ is sought to decrease as the effective temperature is lowered. The latter decreases the capacity of the system to move particles randomly, in other words "thermal" motion counterbalances the weight of the target function S = 1, and decreasing the effective temperature progressively should lead to the best fits. One also needs to keep in mind that if the fit is done on a $S_{exp}(q)$-function defined on a different q-range, the numerical values of $\chi^2$ change. For each simulation run, the q-range is adapted via a change in $q_{min}$ in order to keep the number of particles (and thus the duration of the calculation) reasonable, and therefore minor changes in the initial value of $\chi^2$ are usually found. The latter can be compensated by norming the $\chi^2$-function. During simulated annealing, the algorithm tries to find NP configurations which agree best with the experimental intensity – by keeping the effective temperature fixed, several configurations can be explored for averaging. Note that this process of varying the NP configuration in order to find the experimental scattering is called a reverse Monte Carlo simulation, as opposed to the determination of the hard-sphere structure factor in a direct simulation also used below. In the latter case, the physical input of the simulation is the hard-sphere interaction potential, and not the experimental intensity.

In this article, we will focus on a specific aspect of scattering visible in the intermediate q-range, where the correlation hole and the correlation peak between interacting nanoparticles are normally located. The intensity scattered by the assembly of polydisperse spheres can be calculated using the Debye equation, eq. (1), and the apparent structure factor is obtained by dividing the intensity by the average particle form factor, eq. (2). At low angles, however, the intensity predicted by the Debye equation increases up to the scattering of the entire simulation box – which can be seen as a giant aggregate. It is therefore necessary to calculate the low-q data using a lattice calculation [12, 17, 19-21]. Due to the analytically performed averaging of the Debye equation over all angles, its statistics is better than in the lattice case. Both equations have therefore been combined in all calculations shown here: lattice calculation at low-q, and Debye equation at all higher q. [16, 22]

The comparison to a true hard-sphere fluid is useful to benchmark our calculations, and also as a limiting case for comparison of experimental results. For monodisperse spheres, the easy-to-use Percus-Yevick solution of the Ornstein-Zernike integral equation exists, [23] and several numerical implementations for polydispersity have been proposed in the past. [24-28] Obviously, the PY solution based on excluded volume only, and its polydisperse variant, describe hard-sphere fluids, and not spheres with arbitrary interactions (even if including excluded volume) as studied experimentally by us. It is, however, instructive to compare the predictions of the integral equations to a direct Monte Carlo simulation of particles with only excluded volume interactions. Such simulations have been performed by adapting the RMC code such that $\chi^2$ defined by eq. (3) is ignored. This hard-sphere fluid is simulated with the same particles obeying the same size distribution, with hard-sphere repulsion, in a simulation box of same size. Particle configurations are then equilibrated by random moves taking excluded volume into account. S(q) is calculated by the same subroutine as in RMC, and averaged over hundreds of configurations, until a satisfying statistic is reached.

In Figure 1, an example of a simulation with N = 3392 particles ($L_{box} \approx 440$ nm, $\Phi_{NP}$ = 20%v, $R_0$ = 10 nm, σ = 20%, log-normal size distribution) is exemplarily shown for polydisperse hard spheres. The pure Debye prediction is seen to display the low-q increase discussed above, below q ≈ 0.008 Å$^{-1}$. The simulated composite S(q) is compared to the integral equation prediction for the same parameters – we thank Luc Belloni for sharing his programs with us –, and it is seen to be a trustworthy determination of the apparent structure factor in the entire q-range. This validates our numerical codes.

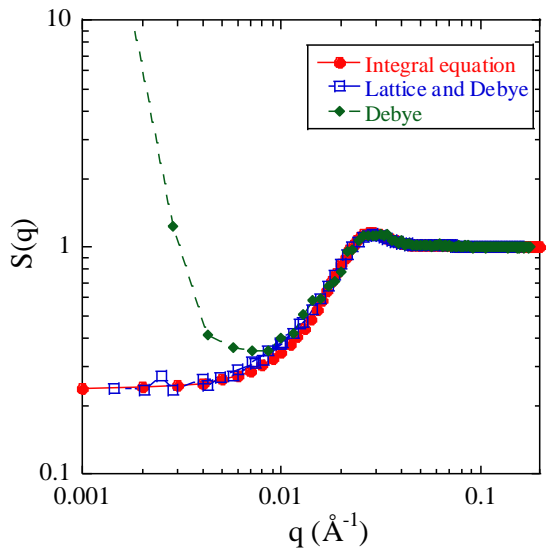

**Figure 1:** Model calculation of apparent structure factor S(q) (squares) calculated for N = 3392 polydisperse spheres ($\Phi_{NP}$ = 20%, $R_0$ = 10 nm, σ = 20%). The Debye prediction (diamonds) showing the low-q increase is superimposed for comparison, as well as the integral equation solution (circles) for spherical particles of the same characteristics.

### 3. Results and discussion

In Figure 2a, the surprising experimentally observed superposition of the scattered intensity of PNCs ($\Phi_{NP} \approx$ 9%v, $R_0$ = 9.6 m, σ = 18%) with the particle form factor P(q) for certain preparation protocols as previously published is shown [13]. Two samples obtained by pre-adsorbing polymer chains in suspension and followed by mixing with the matrix polymer (H10/D10 mixtures, blue symbols; or pure D10, green symbols), and slow drying, show a quite good superposition of the silica intensity with the silica form factor P(q) measured at high dilution in solvent. Comparing eqs (1) and (2), this implies that the corresponding (apparent) structure factor is close to one, i.e., it does not contribute to the scattered intensity. The surprising absence of a structure factor at such high concentrations has been related to an "ideal" dispersion (different from a hard-sphere dispersion, where correlations persist), in analogy with concentrated polymer solutions and melts, where ideal chain form factor scattering may be observed. [13] Our previous explanation, however, remains incomplete, as it is unclear how exactly the hard-sphere repulsion – which is impossible to circumvent for hard and solid silica particles – can be screened. It is the aim of the present article to propose some possible explanations for this puzzling fact.

The structure factor of the same silica NPs is well visible when following a different preparation protocol. In Figure 2a, the scattering of bare NPs embedded in a similar matrix (a blend of H40 and D10, also slow drying, red symbols) is superimposed to the previous data. The correlation hole between q = 0.01 and 0.03 Å$^{-1}$ evidenced by the total intensity passing significantly below the form factor is clearly visible. At high q, the intensity recovers form factor scattering, with possibly a slight structure factor peak at q ≈ 0.033 Å$^{-1}$. This is consistent with a contact distance of 2R$_{NP}$ = 20 nm, i.e., a correlation peak at q = 2π/(20 nm) ≈ 0.033 Å$^{-1}$. Bare NPs thus spontaneously show a structure reminiscent of polydisperse NPs in touch, i.e., an aggregated morphology – also visible from the low-q upturn –, which does not seem to be the case for the pre-adsorbed NPs. The latter seem to be at most weakly aggregated, and even appear to be non-interacting over a large q-range. We will see later that there is a subtle compensation of effects, with an existing but weak low-q upturn and thus attraction. For comparison, we have also performed the calculation of the ideal hard-sphere fluid at the same concentration of particles of the same size and polydispersity. The corresponding intensity (orange line) is seen to deviate from the form factor slightly below 0.02 Å$^{-1}$, and to level off, as expected for particles with only hard-sphere repulsion, and without attractive interactions. None of the samples shown here thus behaves as a true hard-sphere fluid.

The characterization of the degree of aggregation will be important to differentiate "weak" from "strong" aggregation in this article. In previous work, we have made use of the depth of the correlation hole in order to estimate the average density of interacting nanoparticles [19, 29]. The deeper the correlation hole, the higher the local NP density. The example of bare NPs shown in Figure 2a illustrates the presence of a correlation hole (in q-space) corresponding to a local volume fraction of 16%, to be compared to the ca. 10%v nominal one. This may be compared to the densest possible assemblies, which may reach – depending on polydispersity – some 50 or 60%, or more. There are thus locally denser zones in the sample, the dispersion of which may be qualified as still not strongly aggregated. In the present context, however, we will need to differentiate aggregation corresponding to S = 1 with a weak low-q upturn as observed for pre-adsorbed samples in Figure 2a ("weakly aggregated", i.e., presenting a densification), from more aggregated ones, which we thus term "strongly aggregated".

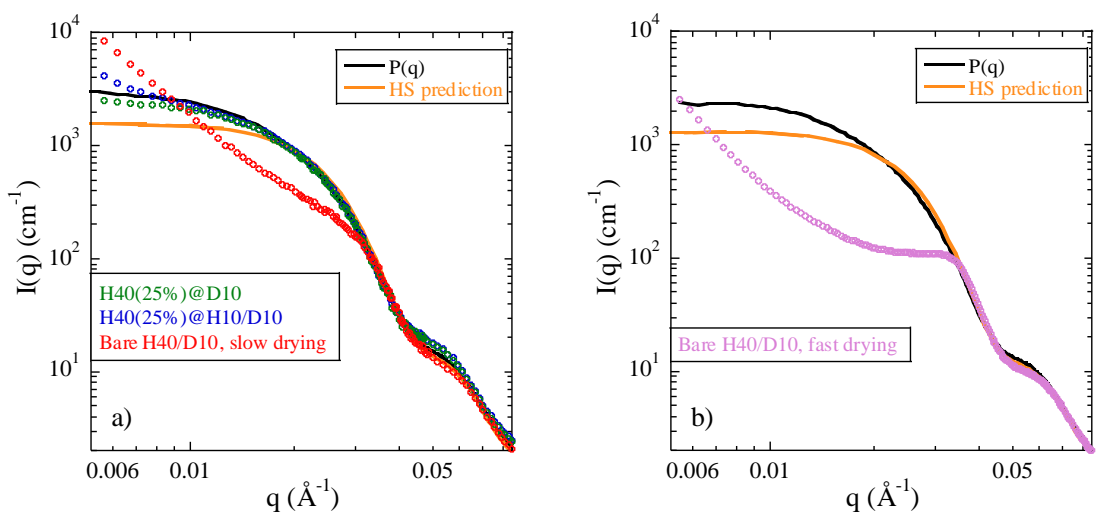

**Figure 2: a)** Scattered SAXS intensity of concentrated nanocomposites with pre-adsorbed NPs (H40(25%)) in PVAc D10 (green symbols, Φ$_{NP}$ = 8.7%v) and in PVAc H10/D10 blend (blue symbols, Φ$_{NP}$ = 9.5%v), and with bare NPs in PVAc H40/D10 blend (red symbols, Φ$_{NP}$ = 8.7%v), all slow drying. The average particle form factor P(q) (black solid line, R$_0$ = 9.6 nm, σ = 18%, measured at 0.7%v in ethanol, rescaled) and the hard-sphere prediction for the same NPs (orange solid line, same concentration) are shown for comparison. **b)** SAXS intensity of a PNC with bare NPs in PVAc H40/D10 blend with fast drying (Φ$_{NP}$ = 8.7%v), compared to its P(q) (R$_0$ = 9.4 nm, σ = 17%) and the corresponding hard-sphere prediction. Adapted with permission from [13]. Copyright 2019 American Chemical Society.

It is possible that the sample casting protocol influences the final nanoparticle configuration. In order to check if the bare NPs initially possess a structure without correlation hole (i.e., S ≈ 1), but lose it during the slow drying process by NP rearrangements, we have accelerated the drying procedure. In Figure 2b, another structure of a PNC made with bare NPs, employing fast drying, is shown and compared to the measured form factor, and the hard-sphere prediction (always done with the same particles at the same concentration). Again, strong particle interaction is observed, confirming that it is the "bare" state of the nanoparticles which is the driving force, and not the speed of drying. In the low-q range, the dominant upturn indicates strong aggregation, which is consistent with the structure factor peak at q ≈ 0.033 Å$^{-1}$. For bare NPs, the structure thus displays neither pure form factor scattering, nor purely repulsive hard-sphere scattering, independently of the drying speed. By comparing Figure 2a to 2b, however, it appears that the slower drying speed results in a more ideal configuration, in the sense that the intensity gets closer to the form factor. Contrarily to our initial intuition outlined at the beginning of this paragraph, one may thus speculate that infinitely slow drying might actually give the time even to bare NPs to arrange in an ideal way (with S = 1), possibly by the progressive establishment of attractive polymer-particle interactions. In presence of pre-adsorbed polymer, this state is spontaneously reached.

As if this were not surprising enough, the situation becomes even stranger when one follows the structure of a pre-adsorbed sample upon annealing. As shown in Figure 3a, the SAXS scattering evolves towards the morphology of strongly aggregated hard spheres, which is recognizable through the correlation hole (the intensity dip at low q) [19], and the considerable low-q upturn. The scattered intensity thus starts from an "unstructured" particle configuration, and as thermal energy is provided over extended times, evolves into strongly aggregated configurations. Again, a purely repulsive hard-sphere structure is never found at intermediate stages, as one can see from the comparison to the hard-sphere prediction (orange lines) in Figure 3a. It is as if the sample had acquired a special configuration of NPs during formulation and drying, apparently caused by the polymer pre-adsorption. This configuration is lost as soon as thermal motion is allowed, and presumably chain desorption takes place leading to bare NPs as observed in Figure 2a. In Figure 3b, the corresponding structure factors are plotted. This representation emphasizes that there is a low-q upturn due to some large-scale aggregation, and a very different behavior for annealed/native samples in the mid-q range. Indeed, below q ≈ 0.03 Å$^{-1}$ (the contact value), the non-annealed PNC shows a S(q) remarkably close to 1 (corresponding to the superposition of I(q) with P(q) in Figure 2a), whereas the same sample after annealing presents a prominent correlation hole. The latter is again evidence for interaction of hard spheres in close contact, and thus for strong NP aggregation triggered by annealing. Confronting this evolution with the structure formed by bare nanoparticles, annealing appears to provide the time and energy to remove the pre-adsorbed chains from the NP surfaces. Finally, at large q, around q ≈ 0.05 – 0.06 Å$^{-1}$, the form factor in polymer is seen to be different from the one measured in a molecular solvent (Figure 2 and 3, before annealing). We have shown recently that this can be traced back to changes in polymer density close to the surface, and this feature disappears with annealing. [13] Due to our data analysis based on the division by P(q) measured in ethanol, this induces artefacts in the same q-range in the structure factor, without being related to interparticle structure. These bumps have thus been disregarded in the present discussion.

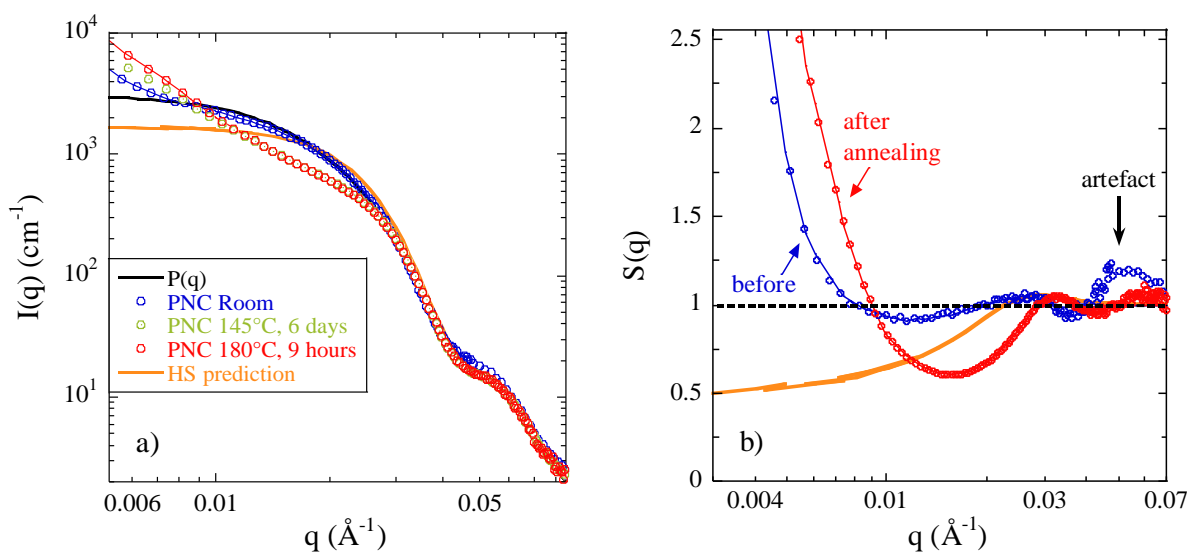

**Figure 3: a)** Scattered SAXS intensity of a concentrated nanocomposite with pre-adsorbed (H40(25%)) NPs in PVAc H10/D10 blend ($\Phi_{NP}$ = 9.5%v, slow drying) measured at room temperature, before (blue symbols) and after subsequent annealing under vacuum as indicated in the legend. The average particle form factor P(q) (black solid line, $R_0$ = 9.6 nm, $\sigma$ = 18%, measured at 0.7%v in ethanol, rescaled) and the hard-sphere prediction for the same NPs (orange line, same concentration) are shown for comparison. Adapted with permission from [13]. Copyright 2019 American Chemical Society. **b)** Structure factors of intensities shown in a), compared to the corresponding hard-sphere prediction. The lines are RMC fits, see text for details. The thin blue and red lines are RMC fits of the same PNC before and after annealing, respectively.

The hypothesis of the present article is to investigate if there may be a particular particle configuration which resembles an ideal one, in the sense that the system scatters as if the particles were penetrable. One of the ideas when starting this study was to see if some special spatial organization of the particle may lead to a smoothing of structure factor extrema. A rather obvious contribution to such an effect could be polydispersity, which leads to the weighting of correlations by different form factors, each with its size and position of oscillations, which might compensate in the final sum in eq. (1). It is thus important to use calculations which include polydispersity by construction.

Reverse Monte Carlo simulations are the perfect tool to study the dispersion of nanoparticles in such samples. In this context, the correct description of polydispersity (as in eq. (1)) proves to be of outmost importance. Before trying to reproduce the experimentally observed intensity in Figure 3 for samples with pre-adsorption, we have performed a variety of test calculations in order to understand which parameters are relevant. To start with, we have calculated the scattering of NP dispersions at a rather high particle volume fraction $\Phi_{NP}$ = 20%v, with log-normal polydispersity in size ($R_0$ = 10 nm) given by $\sigma$ = 1%, 5%, 10%, 15%, 20%, and 40%. We have then used RMC to achieve the best possible agreement with the absence of a structure factor, i.e., defined the target structure factor as $S_{exp}(q)$ = 1. In Figure 4a, the quality of the fit as a function of the different trial steps attempted per particle is plotted for a small system containing approximately N ≈ 1500 particles in the most monodisperse case ($L_{box}$ = 314 nm, the exact number of particles decreases with polydispersity).

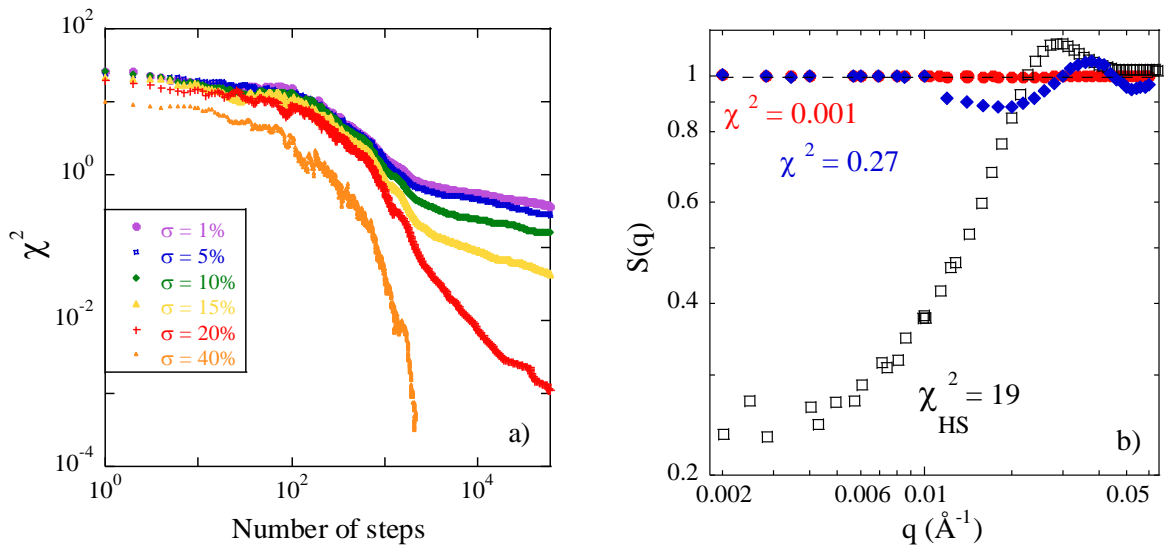

**Figure 4: a)** $\chi^2$ vs number of Monte Carlo trial steps attempted per particle for model calculations varying NP polydispersity ($\Phi_{NP}$ = 20%v, $R_0$ = 10 nm, σ = 1, 5, 10, 15, 20, and 40%, resp.), for small systems. **b)** Best $S(q)$ corresponding to the averaged final NP configurations in a) (filled symbols), compared to the target function $S_{exp}$ = 1 (dashed line), for σ = 20% (plain red symbols, resp. 5%, blue symbols), and to the hard-sphere prediction for the same polydispersity (empty symbols).

The result in Figure 4a illustrates the fact that it is possible to obtain acceptable fits at high polydispersity, but not for too monodisperse particles. Indeed, $\chi^2$ is found to be a decreasing function with simulation time, which however decreases more strongly for higher NP polydispersities. It reaches acceptable level of fit quality of about $\chi^2 \approx 0.001$ only for the highest polydispersity of σ = 20% and 40% – possibly also for 15%, with much longer simulations (not performed). To see the result in terms of fit quality, the best structure factor for σ = 20% is superimposed in Figure 4b to the target function $S_{exp}(q)$ = 1. Only at the very high zoom in y-axis, 1 ± 0.01, differences between the functions would be noted. With σ = 5%, the quality of the fit is seen to be worse. The fit functions are also compared to the usual apparent hard-sphere structure factor for spheres of the same properties (see methods for details) in Figure 4b. The latter function displays a deep correlation hole, down to about $S(q{\rightarrow}0) \approx 0.25$. For comparison, the $\chi^2$-value for this q-range of the hard-sphere prediction compared to $S_{exp}$ = 1 is about 19.

It may be questioned if the small system size of these simulations used in order to obtain quick answers is one of the reasons for the possibility to find solutions with S = 1. In this case, however, it should work for all polydispersities, which is not the case in Figure 4a. Nonetheless, we have increased the system size for a series of model calculations, in spite of the higher costs in simulation time. In Figure 5a, the quality of the fit as a function of the different trial steps attempted per particle is plotted for a system containing approximately N ≈ 4100 particles in the most monodisperse case ($L_{box}$ = 440 nm, the exact number of particles decreases with polydispersity).

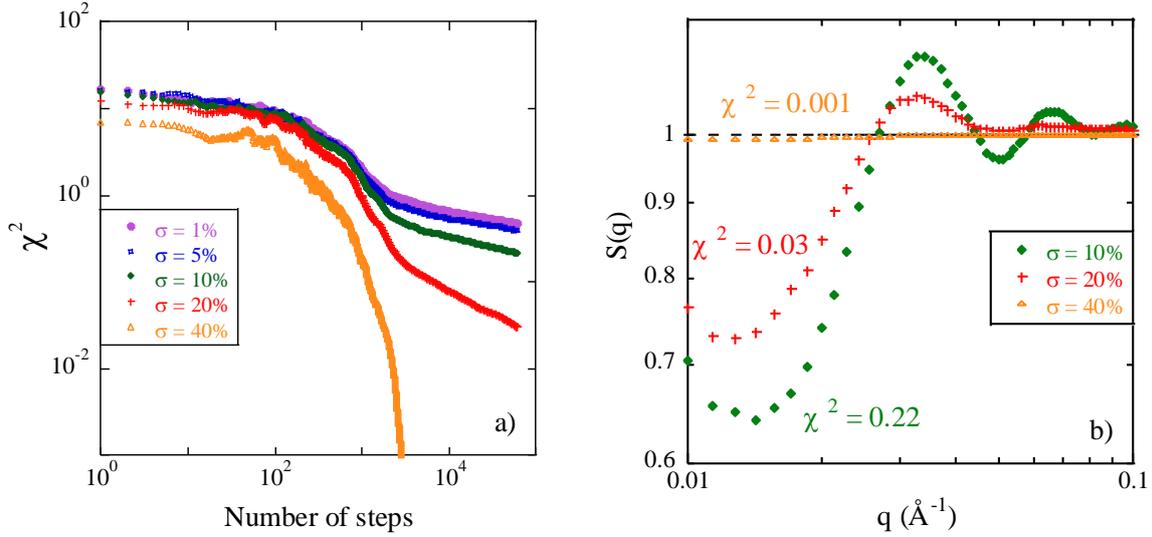

**Figure 5: a)** $\chi^2$ vs number of Monte Carlo trial steps attempted per particle for model calculations varying NP polydispersity ($\Phi_{NP}$ = 20%v, $R_0$ = 10 nm, σ = 1, 5, 10, 15, 20, and 40%, resp.), for bigger systems. **b)** Best $S(q)$ corresponding to the averaged final NP configurations in a), compared to the target function $S_{exp}$ = 1 (dashed line), for σ = 10, 20, and 40%, respectively.

As one can see by comparing Figures 4a and 5a, the bigger systems need considerably more time to converge, and a system with 20% polydispersity which converged well at a small size did not have enough time to find a satisfying solution for more particles. Given the logarithmic time scale, it would last exceedingly long to search for a solution with this system. However, the same tendency as in Figure 4a, higher polydispersity allowing more "ideal" structures, is clearly observed. Indeed, $S(q)$ resulting from the RMC simulation for σ = 40% is seen to be very close to the ideal case $S(q)$ = 1, as these functions are indistinguishable on this scale. If one zooms into the structure in Figure 5b, the function is seen to contain a signal reminiscent of the thermally equilibrated hard-sphere structure shown exemplarily in Figure 4b. This remaining signal becomes weaker with increasing polydispersity. With the speed (and reality) of convergence observed in Figures 4a and 5a, polydispersity is thus found to enable the smearing of the underlying hard-sphere structure. This idea will be further investigated below, in terms of pair-correlation functions. Another interesting feature can be observed in Figure 5b. For all polydispersities, there is a low-q upturn. We have checked that this upturn has a local slope opposite to the one of the form factor of the simulation box, i.e., it is a true indication of structure and is not caused by the finite system size. It seems to indicate that there is some attractive interaction present in the system. We will see below that this is the key to understanding such "ideal" systems.

It appears from Figures 4 and 5 that polydispersity has a dominant role in enabling "ideal" particle structures. One may thus wonder if for a fixed polydispersity, a low enough volume fraction may also open this possibility. Obviously, letting $\Phi_{NP}$ tend to zero is a trivial way of obtaining S = 1, but the question here is if this can be obtained for experimentally relevant volume fractions, where the structure factor is well visible in hard-sphere suspensions. We have therefore performed the same type of RMC simulations, fixing the polydispersity to σ = 10%, and varying the NP volume fraction from $\Phi_{NP}$ = 1 to 25%v. In order not to make simulations at too low particle numbers, the simulation box has been adapted as a function of the volume fractions. As described in the methods sections, this is obtained by modifying the q-range, namely $q_{min}$, yielding a total box size which decreases from $L_{box}$ = 628 nm for the two lowest concentrations, to 440 nm for the intermediate ones, and finally 314 nm for the two highest volume fractions.

As one can read off in Figure 6 from the strong decrease of $\chi^2$ for the lowest volume fractions, it is indeed possible to find ideal particle configurations. There seems to exist a threshold value around $\Phi_{NP}$ = 15%v, above which no ideal solution at this polydispersity is found, at least not on the timescale of simulation. With regards to our experimental results discussed above (Figure 2 and 3a), it is noteworthy that the experimental concentration of ca. 10% is in the range of possibly ideal structure. For higher polydispersities, as shown in Figure 5a, higher concentrations may also allow for ideal structures, and one could thus imagine a "phase diagram" in $\Phi_{NP}$- σ-space, with ideality being situated in the lower concentration/higher polydispersity corner. Two simulations for bigger systems have also been included in Figure 6. They show that at 10% volume fraction there is no problem to find ideal solutions even with bigger systems, whereas it becomes more complicated at 15%.

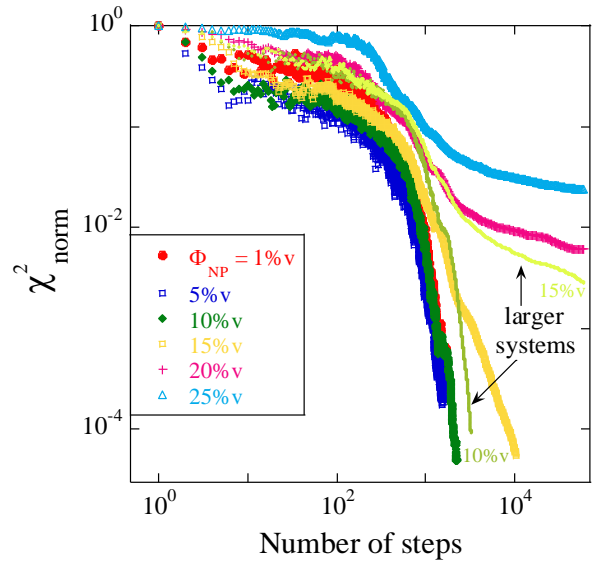

**Figure 6:** $\chi^2$ normed to 1 at the beginning of the simulation vs. number of MC trial steps attempted per particle for model calculations varying NP volume fraction ($R_0$ = 10 nm, σ = 10%, $\Phi_{NP}$ = 1, 5, 10, 15, 20, and 25%v, resp.). The corresponding data as obtained using larger systems made of ca. 2000 to 3000 NPs are included for $\Phi_{NP}$ = 10 and 15%v, respectively.

Now that we have shown that it is possible to find particle configurations by RMC which show close to ideal scattering (i.e., S = 1), for either low concentrations, or high polydispersity, two questions remain to be answered in the framework of the present article. First, what is this special particle configuration in real space, or better, what is its peculiarity in analogy to the observed "ideality" in reciprocal space? And secondly, is it possible to apply this concept to real data as shown in Figure 3? Concerning the first question, we first thought about possible artefacts of the simulation. The simulation box, for instance, might "order" the particles in some specific way, creating density fluctuations capable of compensating the features of the structure factor. We think, however, that this can be ruled out safely. To see this, we have calculated average density profiles for some systems, starting from the box center outwards, and found no evidence for a superstructure. Of course, one might imagine non centro-symmetric density fluctuations, but this idea does not withstand closer examination: if such a superstructure existed, then this mechanism should work for any polydispersity. However, Figures 4a and 5a clearly show that polydispersity is a key parameter, and ideal solutions cannot be found for too monodisperse particle populations.

It is instructive to have a closer look into the particle correlations in reciprocal space in presence of polydispersity. Eq. (1) gives a straightforward access to the partial structure factors, because it is sufficient to choose the spheres entering the double sum in order to obtain the corresponding partial structure factor. This approach allows answering the question if beyond the picture outlined above based on aggregation and polydispersity, there is a special order between, say, bigger and smaller particles. The problem is that the continuous log-normal distribution function makes it difficult to decide who is "big" and who is "small". An obvious approach to simplify the problem of polydispersity is to divide the continuous size distribution in discrete populations, and study each of them separately, as well as their cross-correlations.

We concentrate here on a polydispersity of $\sigma$ = 20%. The corresponding log-normal distribution function of the radii is shown in Figure 7, together with a simple minimal binning in three sizes called "small", "average", and "big". This binning has been defined manually by a symmetric bin positioned at the average radius ($R_{mean} \approx 10.1$ nm) containing ca. 50% of the particles. The bigger bin contains all bigger particles, and their average radius is 13 nm; and the smaller bin contains all smaller ones, of average radius 7.8 nm. Note that this binning as shown in Figure 7 is only a guide to the eye, the bins are positioned on their average radii: binning serves to identify the particles as belonging to a given population, the radius of each particle remains unchanged and still obeys the log-normal distribution (including very small and very big radii, outside of the bins drawn in Figure 7).

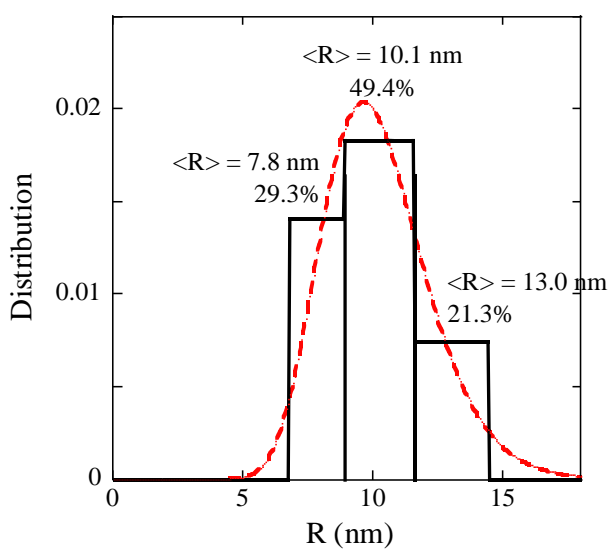

**Figure 7:** Log-normal particle size distribution ($R_0$ = 10 nm, $\sigma$ = 20%, dashed line) with simple binning in "small", "average", and "big" NPs as indicated in the text.

We have calculated the scattering contributions corresponding to each population only, by applying eqs. (1) and (2), and the division of I(q) by P(q), separately to each population of particles. Each corresponds to the result of eq. (1) calculated with the appropriate limits of the double sum, like e.g., for the "small-average" cross-correlation, i = 1 to $N_s$, and j = 1 to $N_{av}$, for the $N_s$ (resp. $N_{av}$) particles in the "small" and "average" bins. Note that this corresponds to the form and structure of each population (and not only the form factor), but the structure factor is necessarily weaker due to the dilution effect of taking only a subset of half or less of the particles. The sum of the three functions does not sum up to the total intensity, as the three cross-correlations "small-average", "average-big", and "small-big", need also to be included.

In Figure 8a, the six functions and their sum are compared to the total intensity for a model calculation with S = 1 ($\Phi_{NP}$ = 20%v, $\sigma$ = 20%, small system). The cross-correlations are seen to be negligibly small.

The striking feature of the partial structure factors in Figure 8a is that they are all flat, meaning that on any scale, the suspension seems to be structureless. This contradicts our first intuitive guess, which made us think that possibly there is a particular order between, say small and big particles, the small ones filling up space between the big ones, and thus accumulating preferentially around them. We will see in direct space, however, that although the signature in q indicates the absence of structure, there is an underlying order which results in flat structure factors.

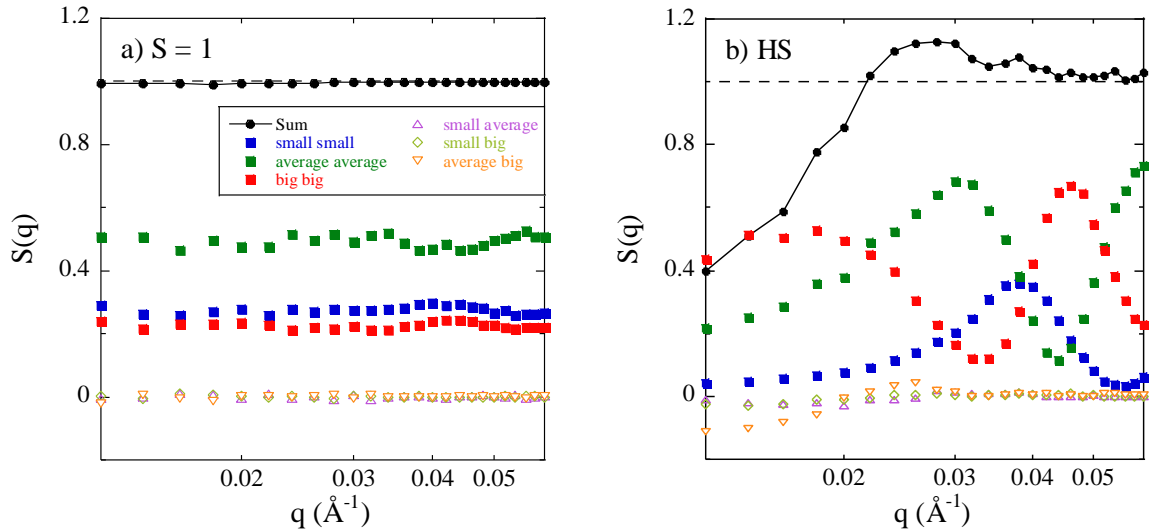

**Figure 8: (a)** Apparent partial structure factors of three populations ("small", "average", "big") for S = 1, $\Phi_{NP}$ = 20%v, σ = 20%, small system). **(b)** Same partial structure factors for an equivalent hard-sphere fluid.

In Figure 8b, the same partial structure factors are plotted for a single configuration of a hard-sphere fluid simply taken from the beginning of the simulation, directly after establishing the initial condition respecting the excluded volume but not the target structure factor S = 1. The sum of the six functions is also shown in Figure 8b, and it is seen to correspond (within statistical error) to the apparent structure factor of hard spheres similar to the one plotted in Figure 1. It is illustrative to compare the shape of the partial structure factors in the case of the hard-sphere fluid (Figure 8b) to the ones with S = 1 (Figure 8a). The former are strongly structured, each presents a maximum at a q vector (q = π/R) corresponding to their typical contact distance, i.e., two radii between centers (binned as shown in Figure 7). The sum of these wildly fluctuating functions is rather well-behaved, i.e., the maxima and minima of the partial functions compensate for all q but for the low-q region, where the correlation hole in q is constructed by the contribution of each sub-population. Needless to say, there is no such thing with the suspensions with S = 1 shown in Figure 8a, which looks completely structureless at all scales. In the latter case, the constant sum of the six partial structure factors is thus not the result of a compensation between population, as it is more or less the case for the hard-sphere fluid beyond the correlation hole. For S = 1, all populations appear to be truly unstructured.

In order to explain the ideal behavior, we had emitted the hypothesis of destructive interferences coming from a special spatial arrangement of particles. We now follow this idea further in real space. The force of the RMC simulation is to provide data on the position of each particle (each with its own size) corresponding to each scattering result, in particular to the best fits. Of course, this is only one possible configuration compatible with the scattering, but during averaging many of them are produced with the same generic features. We have made use of this information in previous work by Musino et al, where the aggregation properties of nanoparticles were analyzed. [12, 17]. Here, we intent to understand how polydispersity can participate in smoothing the intensities so much that even interaction peaks and correlation holes disappear. The pair-correlation function g(r), i.e., the normed probability to find the center of a particle at a distance r from another center, is a useful tool in this respect. For monodisperse spheres, it is zero for center-to-center distances smaller than one NP diameter D due to excluded volume – the region with r < D is termed the correlation hole in real space –, then abruptly raises to the contact value, and levels off at large distances, where the probability to find another sphere is exactly given by the average particle density. Normalizing by the latter quantity thus makes the pair-correlation function tend to one at large distance. In presence of polydispersity, the "diameter" mentioned above is the sum of the two radii of the spheres in question, and this is different for each pair. The resulting pair-distribution function (between particle centers) is then averaged over all pairs of spheres, and the abrupt increase to the contact value is smoothed. Although the structure factor is not the simple Fourier transform of this pair-distribution function for polydisperse systems, one can imagine that this smoothing of the pair-distribution function results in smoothing the scattered intensity. The g(r) is shown in Figure 9 for $\Phi_{NP}$ = 10%v, $\sigma$ = 20%. The system contained ca. 5000 particles ($L_{box}$ = 628 nm), and g(r) corresponds to configurational averages with S(q) = 1. The pair-distribution function is seen to be strongly structured and will be further discussed below. It is also interesting to check the influence of polydispersity, which is the feature which makes the observed structure different from the Fourier transform of the pair-correlation function. While the total g(r) is strongly structured, the partial structure factors are not. We have checked that the partial pair correlations $g_{ij}(r)$ are also peaked, although in a weaker manner, explaining both the flat $S_{ij}(q)$, and, via the sum, the peaked g(r).

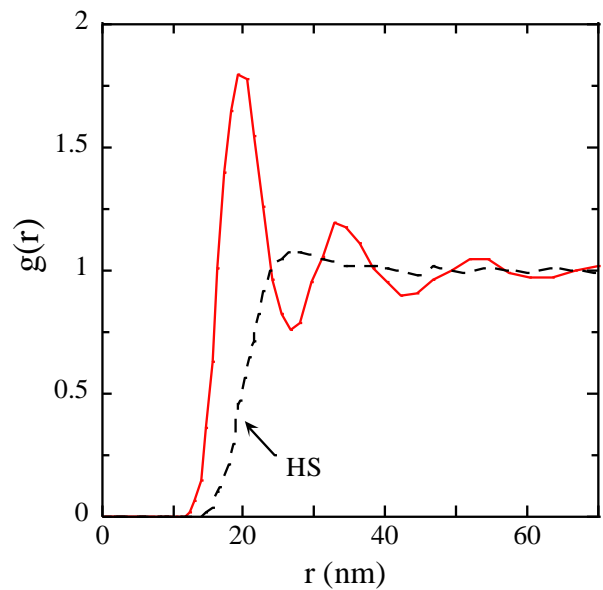

**Figure 9:** Pair-correlation function g(r) extracted from simulation results ($\Phi_{NP}$ = 10%v, $\sigma$ = 20%) after reaching a particle configuration presenting S(q) = 1 (solid line). For comparison, g(r) of the corresponding hard-sphere fluid is also shown (dashed line).

Sticking to the discussion in real space, we now analyze the consequences of aggregation mechanisms. The latter, if not too strong, may lead to S = 1 at least over some length scales and to a low-q upturn as observed in Figures 4b and 5b. The presence of attractive interactions between nanoparticles is very common, as they are often hydrophilic, whereas the polymer is usually hydrophobic. Other types of interactions also exist, e.g., any polymer mediated attraction, like depletion interaction during preparation steps with solvent. As a consequence, locally concentrating particles leads to higher values of the pair-correlation function at the contact value, which is indeed the case as one can see through the existence of a first peak in Figure 9. For comparison, no such strong peak is present in the hard-sphere fluid. Moreover, a second (and also higher order) peak is found; this indicates ordering in close vicinity of particles. The typical distance identified in g(r) is 20 nm, which is in the range of immediate contact between spheres (2R ≈ 20 nm) taking into account polydispersity. One can compare it to the second order closer than twice 20 nm, which is compatible with spheres accumulating in their mutual interstitial sites. Moreover, both of these distances are smaller than the average center-to-center distance between neighboring particles, which would be ca. 37 nm at this concentration for a cubic lattice. Therefore, the g(r) shown in Figure 9 clearly indicates aggregation, i.e., the presence of zones within the suspensions with nanoparticles at a density higher than the nominal (average) density. From the correlation hole analysis mentioned above, the local density of the most aggregated case is not exceedingly large (16%v in the "bare" case), meaning that what we call weak aggregation here corresponds to zones where NPs are closer (i.e., local volume fractions between 10% and 16%) but not necessarily in close contact, with a large range of interparticle distances due to the weak interactions and polydispersity. Obviously, due to volume conservation, there must be also zones of lower density elsewhere in the sample. The close contact between nanoparticles leads to an exacerbation of the (relative) depth of the correlation hole (in r-space, below the diameter), highlighting thus the hard-sphere excluded volume. Intuitively, aggregation alone is thus counterproductive if one seeks reaching S = 1. But in reciprocal space, particle attraction generates the presence of a low-q upturn, the tail of which may replenish the correlation hole. Putting together the two arguments, it thus appears to be possible that the combination of aggregation – with its tendency to emphasize the correlation peak –, and polydispersity – which intrinsically smooths the correlation hole and the peak – might be able of generating "ideal" particle dispersions. If one compares to the "strong" aggregation as encountered, e.g., in the bare system with interparticle attraction mediated by hydrogen bonds in Figure 2, the aggregation must be rather weak, as if the pre-adsorbed protective polymer layer allowed for some weak bridging, or screened van der Waals attraction. As a result, in Figures 4b and 5b, a weak upturn at low-q is found, in agreement with our interpretation of the pair-correlation function.

It is now time to come back to the initial problem of S = 1 encountered in real samples as shown in Figure 3. In this figure, the lines are fit functions produced by reverse Monte Carlo with the experimental apparent structure factor as input, meaning that it is possible to describe these intensities before and after annealing. It has to be noted, however, that it was necessary to extrapolate the intensities to lower angles in order to increase the simulation box size to acceptable values. This extrapolation has been done by respecting the order of the curves, but it obviously contains a part of speculation.

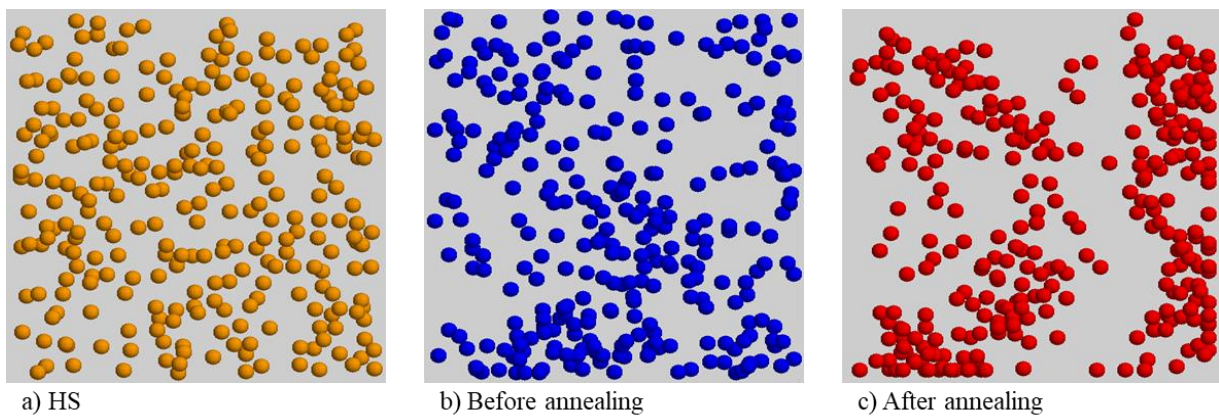

**Figure 10:** Slices of snapshots of the simulation box taken in the center, of thickness 70 nm, for experimental system discussed in Figure 3. Particle polydispersity is present but not represented. **a)** Hard-sphere system taken directly after system construction. **b)** System before annealing. **c)** System after annealing.

In Figure 10, we compare slices in direct space of the simulation box of thickness 70 nm as one would obtain from transmission electron microscopy [12]. The slices are located in the center of the box. The plot program does not reproduce particle polydispersity, but each sphere has a different radius in the simulation. On the left, in Figure 10a, an example of a hard-sphere case is shown. It is seen to be rather homogeneous, and corresponds to the very beginning of the simulation. In Figure 10b, an example of a simulation box satisfying S close to 1 is shown, i.e., it corresponds to the particle configuration fitting the non-annealed intensity in Figure 3a, over the entire experimental q-range plus the extrapolation down to $q_{min}$ = 0.0014 Å$^{-1}$ as stated above. The structure in the slice is still homogeneous, but some holes indicate the presence of denser ("weak aggregation") and less dense zones in the sample. In Figure 10c, finally, the intensity corresponding to the annealed sample is seen to lead to what we call "strong aggregation" in this paper, i.e., there are larger empty zones showing up throughout the sample, and by volume conservation, the remaining zones are thus denser.

As a final result, we note that it is possible to reproduce the experimentally scattered intensities corresponding to S = 1, and that the evolution of the intensities with annealing is the signature of a passage from a state of weak to stronger aggregation. Although it is possible that some ingredient escaped from our analysis, the surprising existence of "ideal" states in rather concentrated particle suspensions can thus be explained by a combination of polydispersity and weak aggregation inducing the existence of weakly correlated, slightly denser zones in the sample.

### 4. Conclusion

The intriguing experimental finding of almost pure form factor scattering of spherical silica beads in rather concentrated polymer nanocomposites after a special preparation protocol involving pre-adsorbed polymer chains [13] has been analyzed by means of reverse Monte Carlo simulations. The principle of reverse Monte Carlo simulations is to seek particle configurations in space with a scattering signature compatible with the experimental data. In the present article, we have used this algorithm to see if model suspensions of polydisperse particles can reach a state of pure form factor scattering, i.e., absence of any apparent structure factor reminiscent of interparticle interferences. We have systematically varied the particle polydispersity and volume fraction, and found that at 10% or 20% volume fraction, a state of "S = 1" can indeed be reached, provided that the polydispersity is high enough.

We have then tried to understand the microstructural state leading to this property. In a first approach, we have calculated the partial structure factors between populations representing small, average, and big particles. These cross-correlations are very strong in the case of concentrated hard-sphere fluids, whereas they become completely "unstructured" ($S_{ij}(q)$ = constant) for S = 1. This implies that there is not a special order of, say, small particles among the big ones, but that on each scale the diversity of radii contributes its share to smoothing the final structure factor obtained by addition of all the partial ones. Reasoning in direct space, i.e., based on the pair-correlation function, shows that polydispersity smooths the edge related to close contact between particles, by filling the correlation hole in real space more efficiently. A direct consequence in reciprocal space is that many particles in close contact correspond to some large-scale aggregation, which are usually identified by low-q upturns, and some indications of which are present both in experimental spectra and in the simulation results. This line of reasoning thus suggests that the combination of some limited densification, which generates a weak low-q upturn, the tail of which fills the correlation hole in q-space, with polydispersity which smooths the correlation hole and the peak, is capable of "simulating" the absence of interactions, i.e., S = 1. Thermodynamically, this seems to correspond to a weak interparticle attraction induced by the presence of the adsorbed polymer layers, possibly screening van der Waals interactions, or allowing for polymer bridges between NPs. Moreover, the response of the system to annealing indicates that the preadsorbed chains may be removed at higher temperatures, thus increasing NP attraction too much and favoring "strong" aggregation incompatible with an ideal state of dispersion.

Besides the specific question raised in the title of this article, this work is intended to shed light on more general issues related to spatial ordering of nanoparticles. First of all, RMC should be seen as a general method to invert scattering data into real space information. RMC has been used in many fields, and for colloidal systems it can be looked at as a statistical method to get average spatial information, which may be as specific as distances between particles, or aggregation numbers. As the method is constructed on the independently measured shape of the nanoparticles, respecting their physical properties like excluded volume, it is a more constrained inversion method as opposed to indirect Fourier transform [1, 2] from intensity to direct space, and thus rather well adapted for suspensions. Its obvious advantage is the possibility to obtain 3D-representations of real samples, and the final result in Figure 10 shows that it helps understanding the possibly complex structure of NP assemblies. RMC is sensitive to the "key features" of the scattered intensity, which makes it a valuable tool, and the resulting configuration files can be seen as a source for further statistical analysis, like the g(r) function, but also generalizations [16, 22]. Also, the structure of particles in nanocomposites is often claimed to be hard sphere-like, but it is unclear to what extent this is true, as such analyses are possibly biased by general fitting approaches with too many free parameters. For instance, particle diameters, polydispersity, and concentration should be fixed independently, by form factor measurements and TGA, in any analysis. Here we show clearly that it is important to benchmark scattered intensities against limiting functions, and hard-sphere structure may be one of them – but not necessarily the only one. We believe that it is important to obtain more quantitative and more trustworthy structures of nanoparticles in space, in particular for nanocomposites where many properties (mechanical, electrical, …) are related to, e.g., aggregation and percolation issues. It is hoped that the discussion of this special case of S = 1 by means of a rather general method of analysis will contribute to any endeavor of understanding nanoparticle dispersions in detail.

## 5. Acknowledgements

L. Belloni (Saclay) is warmly thanked for providing us with a numerical tool implementing a solution of OZ integral equations, allowing the cross check of our results in Figure 1. Fruitful discussions with D. Truzzolillo and M. In (both Montpellier) are gratefully acknowledged. A. Sokolov and V. Bocharova are thanked for setting-up the on-going collaboration, which led to the experimental data reproduced in this article.

## 6. Declarations

The experimental data presented here have been obtained in the framework of an ANR project. We thus acknowledge partial financial funding by the ANR NANODYN project, grant ANR-14-CE22-0001-01 of the French Agence Nationale de la Recherche.

The authors have no competing interests to declare that are relevant to the content of this article.

This article reflects joint work by both authors. Both authors have contributed equally to this article.

Data availability statement:. All experimental data has been previously published, and the relevant simulation results are included in the main body of the present article. If anyone wishes to access position files of particles or similar, we can make them available on (reasonable) request.